\documentclass[prd,aps,floats,twocolumn, tighten]{revtex4}
\usepackage{latexsym}
\begin{document}
\newcommand{\beq}{\begin{equation}}
\newcommand{\eeq}{\end{equation}}
\newcommand{\ie}{{\sl i.e\/}}
\newcommand{\half}{\frac 1 2}
\newcommand{\lag}{\cal L}
\newcommand{\ove}{\overline}
\newcommand{\et}{{\em et al}}
\newcommand{\Prd}{Phys. Rev D}
\newcommand{\Prl}{Phys. Rev. Lett.}
\newcommand{\Plb}{Phys. Lett. B}
\newcommand{\Cqg}{Class. Quantum Grav.}
\newcommand{\Grg}{Grav....}
\newcommand{\Np}{Nuc. Phys.}
\newcommand{\Fp}{Found. Phys.}
\renewcommand{\baselinestretch}{1.2}

\title{The mass of the graviton and the cosmological constant puzzle}

\author{M. Novello}
\affiliation{Instituto de Cosmologia Relatividade e Astrofisica (ICRA-BR/CBPF),\\
 Rua Dr. Xavier Sigaud, 150, CEP 22290-180 \\
  Rio de Janeiro, Brazil}
\date{\today}
\vspace{.5cm}


\begin{abstract}
We propose an interpretation of the cosmological constant puzzle -
i.e., the enormous value of the ratio $\rho_{Pl}/\rho_{vac}
\approx 10^{120} $- in terms of the total number of gravitons in
the observable universe based on a recently discovered
relationship between $\Lambda$ and the mass $m_{g}$ of the
graviton.
\end{abstract}
\date{\today}

\vskip2pc
 \maketitle


\section{Introduction}

It has often been argued that there is no relation between
the cosmological constant and a possible non-null mass of the
graviton.
Such a claim is based on comparing the equation of evolution
of weak metric perturbations satisfying Einstein's first gravitational theory
with the equation of a spin-2 massless
field propagating in the Minkowski background. However, it has
recently been shown \cite{NN1} that this approach relies heavily
on the interpretation of the cosmological constant as a special configuration
states of matter.
This analysis uses Einstein's first description of his gravitational field theory,
hereafter refered to as $EI$, given by the equations of motion
\begin{equation}
R_{\mu\nu} - \frac{1}{2} \, R g_{\mu\nu} = -\kappa \, T_{\mu\nu}.
\label{4Mar1}
\end{equation}
In this framework the cosmological constant is nothing but the energy content
of the matter in a special configuration state called vacuum in which the
corresponding stress-energy tensor takes the form
$T_{\mu\nu} = \Lambda g_{\mu\nu}.$ Here $\Lambda$ should not be understood as a
fundamental quantity of its own, but instead is related to the -
classical or quantum - matter content of the universe.
Our purpose here is not to deal with such an effective $\Lambda$, namely a
complex quantity depending on the presence of matter vacuum.
Instead we are concerned with the bare fundamental cosmological
constant. In the case in which $\Lambda$ is treated as a
fundamental new constant related to the gravitational interaction, one
should adopt Einstein's second description of the equations of
motion, hereafter refered to as $EII$:
\begin{equation}
R_{\mu\nu} - \frac{1}{2} \, R g_{\mu\nu} + \Lambda \, g_{\mu\nu}=
-\kappa \, T_{\mu\nu}. \label{4Mar2}
\end{equation}
Then proving that $\Lambda$ and a non-zero $m_{g}$ are not correlated by
perturbing the Minkowski geometry does not make sense, since  in the absence
of matter $g_{\mu\nu} = \eta_{\mu\nu}$ is not even a solution of equation
(\ref{4Mar2}). Thus the standard argument claiming there is no relation between $\Lambda$ and
$m_{g}$ cannot be used \cite{2}.

In order to analyze the consequences of a similar argument in the realm of
the theory given by eq. (\ref{4Mar2}) one needs to make some important
modifications. The first one is to look for the equivalent fundamental
state of the geometry in this theory, namely the one containing the maximum
number of symmetries generated by the ten Killing vectors. This means dealing
with the deSitter geometry. Thus, one needs to examine the equations
of motion of spin-2 massive and massless fields in such geometry,
and then compare these equations with the evolution of a small perturbation
$\delta g_{\mu\nu}$ of the deSitter geometry under the theory $EII.$
This is done in reference \cite{NN1}. The relevant result is that the equation for the
perturbation $\delta g_{\mu\nu}$ of the deSitter background is
identical to a spin-2 massive field, the mass of which is
proportional to $\sqrt{\Lambda}.$ This property remains valid not
only for deSitter but also for arbitrary background geometries \cite{1}
as long as we stay in the setting of theory $EII.$

\section{Two masses?}
What does one expect the mass associated to $\Lambda$  to be?
Following the standard procedure in the quantum context, the natural quantity
should be constructed using three basic ingredients: $\Lambda$, Planck's
constant $\hbar$ and the light velocity $c.$
We are thus led, up to a numerical factor which, for the time being, we take
of the order of the unity, to the formula
\begin{equation}
m_{g} = \frac{\hbar \, \sqrt{\Lambda}}{c}. \label{4MAr3}
\end{equation}
Using relativity in the quantum world, this is the only way to obtain a formula
for mass using a quantity which has the dimension of length.
Neverthless, as it has often been emphasized, the graviton is not just one of those
particles that happens to exist in any metrical structure: it is
special. This is because the graviton is intrinsically related to
the metrical structure of space-time.
However, one can construct another quantity which has the dimension of mass
using $\Lambda.$ It also contains three basic ingredients, but
exhibits a dependency on the gravity world instead of the quantum world.
This second mass, call it $M_{g}$, is constructed using $\Lambda$, Newton's
constant $G_{N}$ and the light velocity $c.$ This yields, up to a
numerical factor which, for the time being, we take of the order of
the unity, to the expression
\begin{equation}
M_{g} = \frac{c^{2}}{G_{N} \, \sqrt{\Lambda}}. \label{4MAr4}
\end{equation}
Before proceeding let us stop for a while to ponder about the
meaning of this $M_{g}.$ This can be rewritten in an equivalent form as
\begin{equation}
M_{g} = \frac{\Lambda \, c^{4}}{G_{N}} \,
\frac{1}{\sqrt{\Lambda^{3}}} \, \frac{1}{c^{2}}. \label{4MAr5}
\end{equation}
This formula contains three separate terms. The first one
represents the energy density generated by $\Lambda;$  the second
one is the total volume of the universe restricted to its horizon
$(c/H_{o})^{2} \approx \Lambda^{-1}$ and the last term converts
the total energy into a mass. Thus
 we are almost constrained to interpret  $M_{g}$ as the total mass of all existing
gravitons in the observable universe. Writing
\begin{equation}
M_{g} = N_{g} \, m_{g} \label{4Mar6}
\end{equation}
it then follows that $N_{g}$ is nothing but the total number of gravitons
contained inside the observable horizon.

An unexpected result appears when we evaluate this quantity in our
actual universe: it is exactly the same number that appears in the
standard cosmological constant puzzle. Indeed, from the above
expressions we obtain:
\begin{equation}
 N_{g} = \frac{c^{3}}{\hbar \, G_{N}
\, \Lambda} \label{4Mar7bis}
\end{equation}
and for the ratio of the Planck density and the vacuum density one has
\begin{equation}
 \frac{ \rho_{Pl}}{\rho_{vac}} \approx \frac{c^{7}}{\hbar \, G_{N}^{2}}
\, \frac{G_{N}}{c^{4} \, \Lambda}. \label{4Mar7}
\end{equation}
Consequently,
\begin{equation}
\frac{\rho_{Pl}}{\rho_{vac}} \approx N_{g}\approx 10^{120}
 \label{4Mar8}
\end{equation}
This analysis suggests an explanation to the traditional cosmological constant
problem:
the value of $\rho_{Pl}/\rho_{vac}$ is so large because
there is a huge quantity $N_g$ of massive gravitons in the observable
universe, , with $m_{g} \approx\sqrt{\Lambda}.$

\section{Acknowledgements}
 This work was partially supported by {\em Conselho Nacional de
Desenvolvimento Cient\'{\i}fico e Tecnol\'ogico} (CNPq) and {\em
Funda\c{c}\~ao de Amparo \`a Pesquisa do Estado do Rio de Janeiro}
(FAPERJ) of Brazil. I would like to thank Dr Samuel Senti for his
kind help in the final English version of this manuscript.


\begin{references}
\bibitem{1} A comment on this identification is worthwhile.
In the Minkowski geometry the Poincar\'e group of
the symmetries of space-time has two Casimir operators that yields
two label for the associated irreducible unitary representation
(IUR), identified as the mass and the spin. In the case
of the deSitter group $SO(4,1)$ there are also two Casimir operators
but they cannot be identified to the spin and the mass in all cases.
For simplicity and tradition's sake, we still refer to the
deSitter parameters as mass and spin, even when the contraction IUR of the
deSitter group is not a IUR of Poincar\'e group.

Let us also point out that although the mass of the graviton given in equation
(\ref{4MAr3}) is proportional to $\sqrt{\Lambda}$, the corresponding
gravitational perturbation has only two degrees of freedom (see \cite{NN1} for
details).
\end{references}
\end{document}